\documentclass[%
reprint,
groupedaddress,
 amsmath,amssymb,
 aps,
prb,
]{revtex4-1}

\usepackage{graphicx}
\usepackage{dcolumn}
\usepackage{bm}
\usepackage{hyperref}

\begin{document}

\newcommand{\be}{\begin{equation}}
\newcommand{\ee}[1]{\label{#1}\end{equation}}
\newcommand{\bem}{\begin{eqnarray}}
\newcommand{\eem}[1]{\label{#1}\end{eqnarray}}
\newcommand{\eq}[1]{Eq.~(\ref{#1})}
\newcommand{\Eq}[1]{Equation~(\ref{#1})}
\newcommand{\vp}[2]{[\mathbf{#1} \times \mathbf{#2}]}


\title{ Gyrotropic Zener tunneling and nonlinear IV curves in the zero-energy Landau level of graphene in a strong magnetic field }

\author{Antti Laitinen, Manohar Kumar, Pertti Hakonen}
 \affiliation{Aalto University, Low Temperature Laboratory, Department of Applied Physics, Espoo, Finland}
\author{Edouard Sonin}
 \affiliation{Racah Institute of Physics, Hebrew University of
Jerusalem, Givat Ram, Jerusalem 91904, Israel}

\date{\today}

\begin{abstract}
We have investigated tunneling current through a suspended graphene Corbino disk in high magnetic fields at the Dirac point, \textit{i.e.} at filling factor $\nu$ = 0. At the onset of the dielectric breakdown the current through the disk grows exponentially before ohmic behaviour, but in a manner distinct from thermal activation. We find that Zener tunneling between Landau sublevels dominates, facilitated by tilting of the source-drain bias potential. According to our analytic modelling, the Zener tunneling is strongly affected by the gyrotropic force (Lorentz force) due to the high magnetic field.
\end{abstract}

\maketitle


\section{Introduction}
The zero-energy Landau level is a unique feature of a graphene monolayer in a strong magnetic field\cite{ezawa}. Without interaction, the state is four-fold degenerate with respect to two values of true electron spin and two values of pseudospin, which describes distribution of electrons between two valleys in the graphene Brillouin zone. The whole state is half-filled, i.e. one half of possible Landau states are occupied and another half of them are empty (holes). However, Coulomb interaction lifts the degeneracy and the original four-fold degenerate zero-energy states split into two states with a gap between them. All states with the lower energy are now occupied, while all states with higher energy are empty. This leads to the integer quantum Hall effect, and electrons at the lower energy sublevel form an insulator state separated from the higher energy sublevel by an energy gap. It was expected that the insulator state is ferromagnetic \cite{ezawa}, although the experiment by Young et al.\cite{young} and Giesbers et al.\cite{Giesbers2009} did not confirm this. Other options (pseudospin ``ferromagnetism'', e.g.) have also been considered for the ground state\cite{Alicea}.

Independently from the character of the gap, the zero-energy Landau state is an insulator with the gap $\Delta$ of the order of the characteristic Coulomb energy equal to  $e^2/\epsilon \ell_B$ in graphene. Here $\epsilon$ is the dielectric constant, $\ell_B =\sqrt{\Phi_0/B}$ is the magnetic length, $B$ is the magnetic field, and $\Phi_0=hc/e$ is the single-electron flux quantum. But any insulator in a high voltage, which we note by $V_{cr}$, becomes a conductor (dielectric breakdown). At bias voltages $V$ exceeding $V_{cr}$, i.e. after the dielectric breakdown, the $IV$ curve is close to linear (ohmic regime). On the other hand, at electric fields essentially less than $V_{cr}$ the conductance is very small and strongly nonlinear. At low temperatures, an exponentially small current $I$ emerges due to Zener tunneling between the two bands \cite{Zener1934}, creating an electron in the empty upper band and a hole in the full lower band. \cite{ziman} Without a magnetic field
\be
  I\propto e ^{-V_Z/ V },
     \ee{z}
where
\be
V_Z \sim {\Delta^{3/2} d \over e \sqrt{E_b} a}
         \ee{}
must be close to $V_{cr}$. Here $a$ is the lattice constant, $d$ is the distance between contacts, and $E_b$ is the band width.

In the quantum Hall state the two bands are flat in the bulk and hence the tunneling current vanishes.  In a strong magnetic field the derivation, which leads to \eq{z} becomes invalid. Motion of electrons in a strong magnetic field is not determined by the Newton's second law with the inertial force proportional to the electron mass, but by the equation of motion of guided centra (centra of Larmor circles around which electrons move). In this equation the external force on the electron is balanced by the Lorentz force and the inertia can be neglected. The equation of motion of guided centra is similar to the equation of motion of quantized vortices in superfluids and clean superconductors with the Magnus force (analog of the Lorentz force on an electron) balancing the external force.

Vortex-like equations of motion essentially modify the semiclassical theory of quantum tunneling. For vortices it was first demonstrated by \citet{Vol72}, who considered nucleation of a circular vortex half-loop near a plane boundary. Probability of quantum tunneling derived from the semiclassical theory based on the  equations  of vortex motion was confirmed by the results of the many-body approach based on the Gross--Pitaevskii equation.\cite{Son73} It is possible to demonstrate\cite{EBS} how the theory of usual quantum tunneling for a massive electron governed by Newton's law transforms to the theory based on the equation of vortex motion when the inertial force proportional to the electron mass becomes much weaker than the gyrotropic force (Magnus force on the vortex, or the Lorentz force on electrons in a strong magnetic field).

The important role of Zener tunneling on electron transport in the quantum Hall regime of a 2D electron gas has already been discussed in the context of quasi- elastic-inter-Landau-level scattering (QUILLS) \cite{Heinonen1984,Eaves1984} and in the connection with magnetoresistance oscillations in the ohmic regime\cite{zener,Bykov2012}. Here we present direct measurement of a current produced by Zener tunneling and calculation of its probability in the subohmic regime, before the breakdown. We analyze Zener tunneling between two Landau sublevels emerging from the zero-energy Landau level (around filling factor $\nu = \frac{hn}{eB} = 0$, where $n$ is the charge carrier density) using the semiclassical theory of tunneling. Also, we measured nonlinear $IV$ curves experimentally at low voltages (weak electric fields) and argue that they provide evidence of Zener tunneling governed by vortex-like equations of motion. This type of quantum tunneling was called Hall tunneling for vortices in superconductors\cite{BlatRev}  and Magnus tunneling for vortices in superfluids\cite{EBS}. Here we call it gyrotropic tunneling because of the crucial role the gyroscopic force (Lorentz force on electrons or Magnus force on vortices) in the process of tunneling.

The $IV$ measurements were done on undoped suspended graphene in the Corbino disk geometry. Corbino geometry has a distinct advantage over Hall bar geometry: no edge states persist and the breakdown takes place through the states in the bulk. Consequently, this geometry has been successfully employed for studying phenomena related to transitions between Landau levels in 2D electron gases: microwave-induced resistance oscillations \& zero-resistance states \cite{Yang2003}, and phonon-induced resistance oscillations \cite{Liu2014}. It has also been employed in studies of edge channel tunneling \cite{Liu1998}, regular Zener tunneling between different Landau-levels \cite{Goran2013}, as well as the bootstrap electron heating (BSEH) model \cite{Komiyama1985} leading to breakdown of the quantum Hall effect \cite{Ebert1983,Cage1983,Nachtwei1999,Komiyama2000a}.

By fitting theoretical formulas to the experimentally obtained $IV$ curves, we could reveal the interval of voltages where the measured current is reasonably well described by the exponential law
\be
  I\propto e ^{-(V_Z/V)^2 },
     \ee{vor}
which follows from the gyrotropic Zener tunneling theory in a strong magnetic field.  This provides evidence of quantum tunneling processes governed not by the particle mass but by the gyrotropic force on a particle.

\section{Theoretical results}

 Since we address the semiclassical theory of quantum tunneling, we need the classical equations of gyrotropic motion for guided centers (in cgs units):
\be
{e\over c}[\dot{\bm r} \times \bm B] = -  {\partial {\cal E} \over \partial \bm r},
    \ee{}
where ${\cal E}$ is the energy of the electron with the 2D position vector $\bm r$. The right-hand side of the equation is the external force on an electron and the left-hand side is the Lorentz force. This vector equation is equivalent to two
 Hamiltonian equations,
\be
\dot x = {\partial {\cal E} \over \partial P_x} ,~~\dot P_x = -  {\partial {\cal E} \over \partial x},
   \ee{}
    for the pair of canonically conjugate variables $x$ - $P_x$, where the conjugate momentum is connected with the second coordinate $y$ of the guided center:
\be
P_x =-{eB\over c}y.
      \ee{}

\begin{figure}[b]
\includegraphics[width=.5\textwidth]{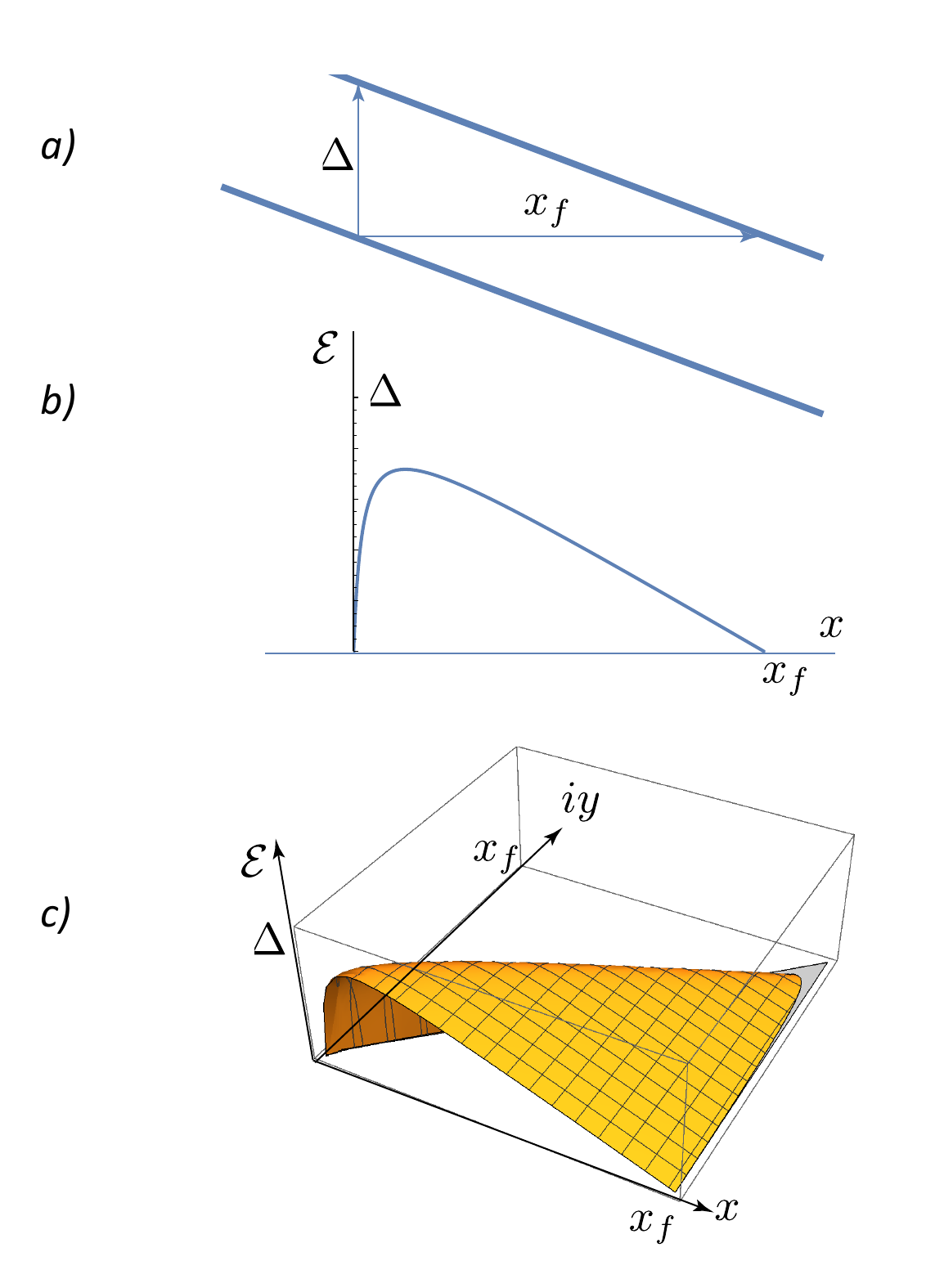}
\caption[]{Zener tunneling between two Landau sublevels (flat bands).
a) Two bands with the gap $\Delta$ tilted in a weak electric field.
b) Plot the energy $\cal E$ vs. the real coordinate $x$ in the interband space.
c) Analytical continuation of the energy as a function of coordinates from the  real plane $y=0$ to the complex plane ${\cal E}=0$.}
\label{Fig1}
\end{figure}

We consider tunneling between two bands separated by the gap $\Delta$. In an electric field the bands are tilted (Fig.~\ref{Fig1}a), and the energy of the electron in the upper  band is equal to the electron energy in the lower band at the distance
\be
x_f ={\Delta\over eE}.
     \ee{x0}
 An electron  tunneling from the lower band to the upper band leaves in the valence band a hole. The electric field $E = V/d$ 
 drives the electron and the hole in opposite directions but this is resisted by the Coulomb attraction between the electron and the hole.  The energy of the electron is
\be
{\cal E}=\Delta -{e^2 \over \epsilon r} -e\bm E \cdot \bm r.
        \ee{bar}
The position vector $\bm r$ connects the positions of the electron and the hole. For the sake of simplicity we assume that the hole is at rest, and only the electron participates in the tunneling event. We consider a weak electric field parallel to the $x-$axis. The shortest path across the interband barrier is at $y=0$ when (Fig.~\ref{Fig1}b)
\be
{\cal E}=\Delta -{e^2 \over \epsilon x} -eEx =\Delta\left(1 -{\ell_B \over  x} -{x\over x_f}\right).
        \ee{barX}
At a weak electric field, $x_f \gg \ell_B$, and two zeroes of the energy are very close to the points $x=0,~y=0$ and $x=x_f,~y=0$:
\be
x_1 \approx \ell_B,~~x_2 \approx x_f-\ell_B.
    \ee{}
 One should remember that we treat electrons as point-like objects, although any electron is distributed in space over the scale $\ell_B$. Thus, an electron position can be determined only with uncertainty $\ell_B$. Therefore differences between $x_1$ and $x=0$ and   between $x_2$ and $x=x_f$ are   within the error bar of our analysis.

In the semiclassical theory of quantum tunneling  one must  find a trajectory of the electron at constant energy (equal to 0 in our case). Such a trajectory is possible only for imaginary $y$ determined from \eq{bar} at ${\cal E}=0$:
\bem
y =-i \sqrt{x^2-{\ell_B^2x_f^2\over (x_f -x)^2}}.
      \eem{}
Analytical continuation of the energy as a function of coordinates from the  real plane $y=0$ to the complex plane ${\cal E}=0$ is shown in Fig.~\ref{Fig1}c.
Since $x_f \gg \ell_B$, it is accurate enough to approximate the trajectory in the complex plane by a simple linear function $y \approx -ix$.

After the trajectory has been found we can calculate the exponent in the exponential law for a small current:
\be
I \propto e^ {-\Gamma}, ~~\Gamma={2\mbox{Im} S\over \hbar}.
       \ee{}
Here $\mbox{Im} S$ is the imaginary part of the classical action variation along the trajectory:
\be
S=\int_0^{x_f} P_x dx=  { eB\over c}\int_0^{x_f} ix \,dx=i{ eBx_f^2\over 2c}.
      \ee{}
Thus, the exponent is
\be
\Gamma ={2 \pi Bx_f^2\over \Phi_0}={2 \pi B\Delta ^2\over \Phi_0 e^2 E^2}.
       \ee{gamma}
The exponent of the law is of the order of the number of flux quanta in the area $\sim x_f^2$, while for vortex tunneling the exponent is the number of bosons in the same area.\cite{EBS}  Remarkably, the exponent depends only on the thickness of the barrier but not on its height. Qualitatively it is of the order $x_f^2/\ell_B^2$, i.e. of the order of the exponent of the overlapping integral for wave functions of two Landau electron states at the distance $x_f$ between them.

\Eq{gamma} is exact if we know the gap $\Delta$. However, we know it only by the order of magnitude:
$\Delta \sim e^2/\epsilon \ell_B$. This yields the exponential law (\ref{vor}) for the current with the characteristic electric field:
\be
V_Z \sim {\sqrt{2 \pi}e B \over \epsilon \Phi_0}d \sim {e  \over \epsilon \ell_B^2}d.
      \ee{ZenerField}
The derived exponential law is valid when the voltage $V$ is much smaller than $V_Z$. At $V \sim V_Z$, the exponential growth of the current saturates, and a crossover to the ohmic regime on the $IV$ curve takes place. Our theory addresses only the exponential law of the tunneling probability ignoring the prefactor. The calculation of the prefactor is much more difficult and would require a lot of information on scattering processes and on spin (pseudospin) structure of electron-hole pairs.

\begin{figure}[h]
\includegraphics[width=.45\textwidth]{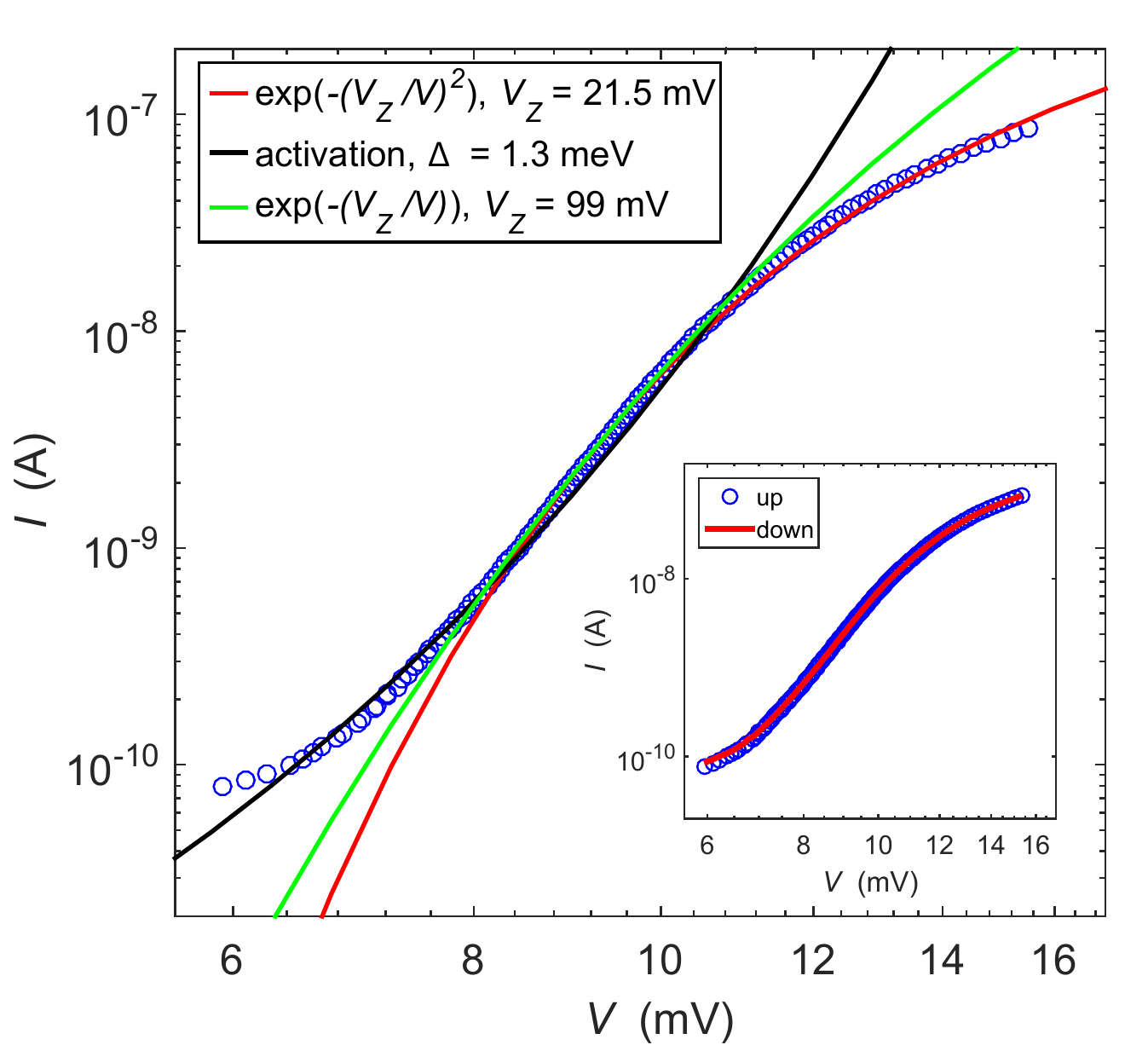}
\caption[]{$IV$ curve of sample S2 at a magnetic field of 2 T using positive bias on the inner contact. Rings are data points, and solid lines denote theoretical curves for gyrotropic Zener tunneling (red) and "normal" Zener tunneling (green). The black solid line represents the activation model of Eq. \ref{activation2} using $N$ = 6. The inset shows the same IV as in the main picture when ramping the bias voltage up (blue circles) and down (red solid line).}
\label{Fig3}
\end{figure}

\section{Experimental results}
The $IV$ characteristics measured at $B$ = 2 T using positive bias on the inner Corbino contact are presented in Fig. \ref{Fig3}. The data are compared with two different theoretical models based on Zener tunneling: the conventional model $\propto\exp(-(V_{Z}/V))$ of Eq. \ref{z}, and the gyrotropic model of Eq. \ref{vor}. The third curve is based on thermal activation, where the number of independent activated regions is a fitting parameter (see below). The data presented in this section is from our sample S2 (see Methods section); very similar results were found on sample S1.

In the case of Zener tunneling, both models can be fitted to the data but the gyrotropic Zener tunneling yields a better agreement. More importantly, the fitting parameter $V_{Z}$ finds more reasonable values when using the gyrotropic model. By reasonable we refer to the fact that the characteristic voltage $V_Z$ should be close to the critical voltage $V_{cr}$, the point where significant current starts to flow through the bulk of the Corbino disk. We define the critical voltage $V_{cr}$ = $V$($I$ = 10 nA), which correspond to the voltage in the middle of the region where the current grows significantly. One can note that $V_{cr} \approx$ 10 mV in Fig. \ref{Fig3}. By comparing this to the values of $V_Z$ in Fig. \ref{Fig3}, 21.5 meV for the gyrotropic tunneling and 99 meV for the conventional case, along with the better agreement of the fit, we conclude that the gyrotropic model is the best to account for our observations at $B$ = 2 T. Moreover, the same conclusion is found at negative bias voltages.

In addition to the Zener tunneling models, we have also considered conduction by thermal activation across localized states. In Fig. \ref{Fig3}, there is a fit based on such an activation model with tunneling current between impurity islands given by:
\be
	I = I_0\left[ e^{-\frac{\bar{\Delta} - eV/N}{k_BT}} - e^{-\frac{\bar{\Delta} + eV/N}{k_BT}} \right],
		\ee{activation2}
where $N$ is the number of islands in the percolation chain, $\bar{\Delta}$ is the activation gap energy, and $I_0$ is a prefactor. By assuming that the tunneling happens through a series of equidistant jumps, the effective voltage is reduced by a factor of $N$, which in turn makes the $dI/dV$ slope shallower. Without setting $N$ = 5 - 10, the activation models become too steep to fit the data. Additionally, this model requires us to set the effective temperature to $T$ = 1.7 K and the gap energy down to $\bar{\Delta}=1.3$ meV (compared to the measured $eV_ {cr} \approx 10$ meV from the critical voltage). Even with these assumptions in the activation model the gyrotropic model nevertheless results in better fits in the region where the dielectric break down takes place ($V \sim V_{cr}$).

$IV$ curves were also investigated at different magnetic fields, $B$ = 1 - 9 T, and fits such as the ones in Fig. \ref{Fig3} were made in order to extract the magnetic field dependence of $V_Z$ and $V_{cr}$. $V_Z$ was obtained from the gyrotropic Zener tunneling fits, whereas the critical voltage $V_{cr}$ was taken as the voltage needed to drive $I = $ 10 nA through the device, as defined before.
\begin{figure}[t]
\includegraphics[width=.45\textwidth]{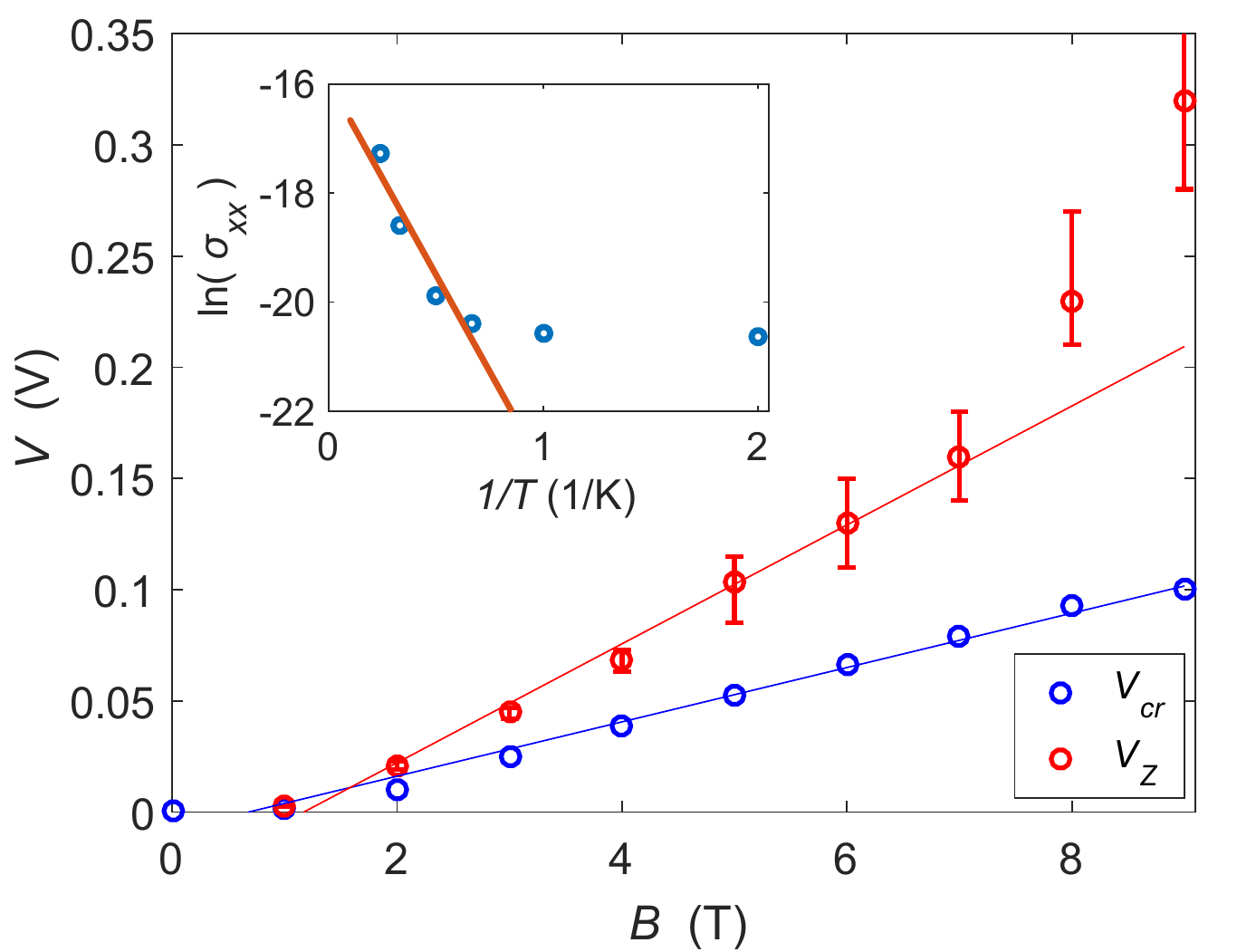}
\caption[]{Dependence of $V_{cr}$ and $V_Z$ vs. $B$, where the latter was extracted by fitting the gyrotropic Zener tunneling model $I \propto \exp{(-(E_Z/E)^2 })$ at different fields to the positive bias data (sample S2). Rings are data points, and the lines are linear fits (two high field points have been omitted in the case of $V_Z$). The dependences are close to linear, and the values of $V_Z$ and $V_{cr}$ are close to each another as expected from the theory. The inset displays an Arrhenius plot from sample S1 at $B$ = 1.8 T. Blue circles are data at $T$ = [0.5, 1.0, 1.5, 2.0, 3.0, 4.2] K and the red line denotes the Arrhenius fit $\sigma = \sigma_0 \cdot \exp{(-\Delta_0/2k_BT)}$ to the four highest temperature points. The gap for $\nu = 0$ obtained from the fit was $\Delta_0$ = 1.2 meV.}
\label{Fig5}
\end{figure}

The extracted $V_Z$ and $V_{cr}$ are displayed in Fig. \ref{Fig5} for the positive bias data. It can be seen that an approximate correspondence $V_{cr} \sim V_Z$ is valid over the whole range as well as approximately linear magnetic field dependence in accordance with Eq. (\ref{ZenerField}). There is a factor of two difference between $V_Z$ and $V_{cr}$ throughout the magnetic field range, which was already noted in the data at $B$ = 2 T in Fig. \ref{Fig3}. The conventional Zener tunneling yields $V_Z$ that is about ten times larger than $V_{cr}$. A theoretical estimate for $V_Z$ can be calculated using Eq. (\ref{gamma}), converted to SI units $V_Z = \frac{eBd}{\sqrt{8\pi}\epsilon \Phi_0}$, where the Corbino disk ring width $d = (D_0 - D_i)/2$ (using the definitions of the Methods section), and $\epsilon = \epsilon_0\epsilon_r$ the permittivity of graphene. The effective value of relative permittivity $\epsilon_r$ = 5 was used\cite{Alicea}. The model yields the observed linear behaviour, $V_Z \propto B$, but with $\simeq 10$ times higher theoretical threshold values calculated from the equation above for $V_Z$. Since the measured $V_Z$ are reasonable in the sense that they correspond to $V_{cr}$, there must be a substantial inaccuracy in the parameters employed in the theoretical estimation of $V_Z$. The most likely culprit is the estimate for the gap $\Delta \sim \frac{e^2}{\epsilon \ell_B}$ that is poorly known and sample dependent\cite{young, Abanin2013}.

\section{Discussion}
There have been several investigations on electric breakdown in QH systems. Various models have been developed, which in addition to Zener tunneling also include thermal instabilities. Thermal runaway, i.e. positive feedback where increase in input power increases conductance that in turn increases power dissipation, would be one option \cite{Komiyama1985}. The critical field $E_c$ has been found to scale approximately linearly with the magnetic field and the largest electric field reported for $E_c$ amounts to $\simeq 25$ kV/m at 9 T  \cite{Yokoi1998} (see the analysis and compilation in Ref. \onlinecite{Komiyama2000a} for filling factors $\nu= 2,4,6$). Our 9 T result at $\nu=0$, $E_c = 100$ kV/m, is by a factor of four larger than the quoted result in GaAs heterostructures. On the basis of inelastic electron - acoustic phonon scattering, the BSEH model leads to $E_c \propto B^{3/2}$, which has been verified experimentally \cite{Hata2016}, but even in these experiments $E_c$ extrapolates only to 14 kV/m at 9 T. Taking also into account the weak coupling to acoustic phonons in similar graphene samples \cite{Laitinen2014}, we conclude that in graphene we are not dealing with thermal runaway at the onset of the nonlinear IV characteristics. Further evidence of nonthermal origin of the IV characteristics is provided by the absence of hysteresis, which one would expect to exist for this type of instabilities. The inset in Fig. \ref{Fig3} displays IVs when ramping the bias voltage up (blue) and down (red), and subsequently the curves laying on top of each other indicates negligible hysteresis. One factor which may contribute to the absence of the hysteresis is the equilibration of the edge states at the metal/graphene contacts in Corbino samples\cite{Slobodeniuk2013} and the shortness of the bulk graphene: the avalanche type of excitation generation, typically assumed to be present in BSEH analysis, requires a substantial distance which can be even in excess of 100 $\mu$m \cite{Komiyama1996,Woodside2001}. Note that BSEH type of behavior with $E_c \propto B^{3/2}$ has been observed in epitaxial graphene with sample lengths $5-35$ $\mu$m \cite{Webber2013}.

In Fig. \ref{Fig3}, we compared our data against two quantum tunneling models based on Zener tunneling: the conventional form and the gyrotropic model presented here. Evidently, the fit to the law of gyrotropic tunneling looks better over the full current variation by nearly three orders of magnitude. For the gyrotropic tunneling, moreover, the fitting parameter $V_Z$ is closer to the critical voltage $V_{cr}$ than for the usual Zener tunneling. According to the theory of Zener tunneling, the critical voltage $V_{cr}$ should be on the same order of magnitude as the voltage scale of tunneling $V_Z$. This relation is clearly better valid for the gyrotropic tunneling than for the regular Zener tunneling. Additionally, we observe linear field dependence of $V_Z$ and $V_{cr}$ in the range $B$ = 1 - 7 T, which is characteristic for the gyrotropic Zener tunneling but not for the conventional model. While our observations seem to agree well with the presented model for the gyrotropic Zener tunneling, the measured value of $V_Z$ is smaller by a factor of ten compared to that obtained from the order of magnitude estimation in Eq. \ref{ZenerField} using an energy gap of  $\Delta \sim e^2/\epsilon \ell_B$.

We assign this discrepancy between the gyrotropic theory and our experiment to the uncertainty in the effective gap value in the transport experiment at zero filling factor. In similar context, observation by the Singh and Deshmukh {\it et al.}~\cite{Vibhor2009} shows discrepancy in the observed activation gap energy and the breakdown Hall voltage. Moreover, if we compare the transport result $\Delta \simeq 10$ meV of Ref. \onlinecite{young} with the result $\Delta =58$ meV of the scanning SET measurement of Ref. \onlinecite{Abanin2013a} (both at at $B=9$ T), we note a difference by a factor of six. Note also that our Arrhenius data in Fig. \ref{Fig5} inset scaled up to $B=9$ T (with linear field dependence) corresponds to $\Delta \simeq 7$ meV. Taking this effective gap reduction factor into account, we find that our theory and experiment agree within a factor of two when using our measured $\Delta$ in Eq. \ref{gamma} instead of $\Delta \sim \frac{e^2}{\epsilon \ell_B}$.

Reduction of the gap might be induced by the nonuniformity of the charge density caused by the bias voltage. In our experiment at low bias, the charge density is not altered by the source-drain biasing, and the measured state will settle to the $\nu$ = 0 quantum Hall state with a robust gap \cite{young}. The relatively high asymmetric source-drain bias offsets the chemical potential in a manner dependent on the transport path and contact resistances (possible with Schottky barriers, see below). The asymmetric bias is due to the use of a low-noise transconductance amplifier, which acts as a virtual ground for the  terminal connected to it. As a consequence, there is an increase in the charge density near the biased lead, which will reduce the energy separation with respect to the next Landau sublevel. Since the other end is connected to virtual ground, there is going to be a region where the gap is unmodified. We expect that the transport on the excited Landau level is similar to the lower level which guarantees electron-hole recombination while the charge is transported across the disk. Hence, the region with large gap is expected to be the dominant one in the experiment, and our results correspond to the regime close to the Dirac point with the filling factor $\nu$ = 0. This conclusion was checked at $B=2$ T by measuring the same IV characteristics in a symmetrized biasing configuration, where the sample was biased on both sides using current biases that resulted in chemical potential offset $\Delta \mu = \pm eV/2$ on the opposite ends. This way, the chemical potential remained at $\mu = 0$ in the middle of the graphene ring, and the tunneling must have happened through the strongly gapped region.

At much higher fields, the cyclotron radius of charged particles is greatly reduced and the carriers can be treated as point-like objects. The breakdown mechanism will be dominated by the electric-field-induced local breakdown within the localized states in the bulk, leading to an avalanche type of breakdown\cite{Komiyama2000a}. Large Fano factors, observed in recent noise measurements on Corbino disks \cite{Chida2014,Hata2016}, have yielded support to the avalanche picture of electron transport at the breakdown point of the quantum Hall effect in GaAs heterostructures. It seems that our IV results also include similar avalanche-induced transport at large magnetic fields \cite{AvalanchePaper2018} where the current increase in the IV curves is actually faster than predicted by our gyrotropic tunneling model. Such enhanced current by avalanches could explain the deviation of the observed $V_Z$ values in excess of the linear behavior in Fig. \ref{Fig5}.

Fermi level pinning of metallic contacts may also play a role in the gapped transport phenomena. In semiconducting CNTs e.g., this leads to FET operation that is based on Schottky barrier modulation \cite{Jimenez2006}. In our devices this effect may come into play because the charge density in graphene is always nonuniform due to the contacts inducing $n$-type doping. The charge doping by contacts leads to $pn$ interfaces at negative bias voltages \cite{Laitinen2016}. In the presence of the magnetic-field-induced energy gap, the nonuniform doping may lead to Schottky-barrier-like structures that would have influence on the Zener tunneling phenomena. Consequently, we have concentrated on the Zener tunneling results on the positive bias where the sample is unipolar.

In conclusion, we have measured tunneling current through suspended graphene Corbino rings in the middle of the zero-energy Landau level corresponding to the filling factor $\nu$ = 0. We found that the tunneling current near the dielectric breakdown is consistent with a nonstandard quantum tunneling process, namely the gyrotropic Zener tunneling where the quantum tunneling takes place along a curved path due to the presence of a large magnetic field. The gyrotropic tunneling model also accounts for the observed linear increase of $V_Z$ and $V_{cr}$ as a function of magnetic field. The presented model fits the qualitative features of our data well by reproducing the exponential growth of current with a characteristic voltage $V_Z$ of expected magnitude in relation to the onset of current. This agreement in the magnitudes of the measured activation gap energy and $V_Z$, calls attention to the $I V$ characteristics as an alternative approach for Arrhenius type measurements. We also note that the $\nu = 1 $ state supports skyrmionic excitations \cite{Shkolnikov2005,Lian2016}; this gyrotropic Zener tunneling model could be a promising tool to investigate such excitations.

\section{Methods}
The Corbino disk geometry has the unique trait that there are no edge states connecting the measurement ports of the device at high magnetic fields, unlike in Hall bar devices. This allows direct probing of the bulk and dielectric break down of the system in the Corbino disk. We study this break down in current annealed suspended graphene devices.

The samples used in this work were fabricated in a manner described in Ref. \cite{Kumar2016}. The fabrication was based on a sacrificial layer of lift-off-resist (LOR) which acted as support for our leads. The LOR was spun on a SiO$_2$ covered p++ doped silicon substrate that also served as a back gate. The graphene part, and its immediate surroundings, were suspended by exposing the LOR in that area to e-beam and dissolving the exposed parts in ethyl lactate, and then rinsing in hexane where low surface tension allowed us to simply lift the chip out of the liquid without destroying the delicate suspended graphene membrane.

\begin{figure}[t]
\begin{center}
\includegraphics[width=.45\textwidth]{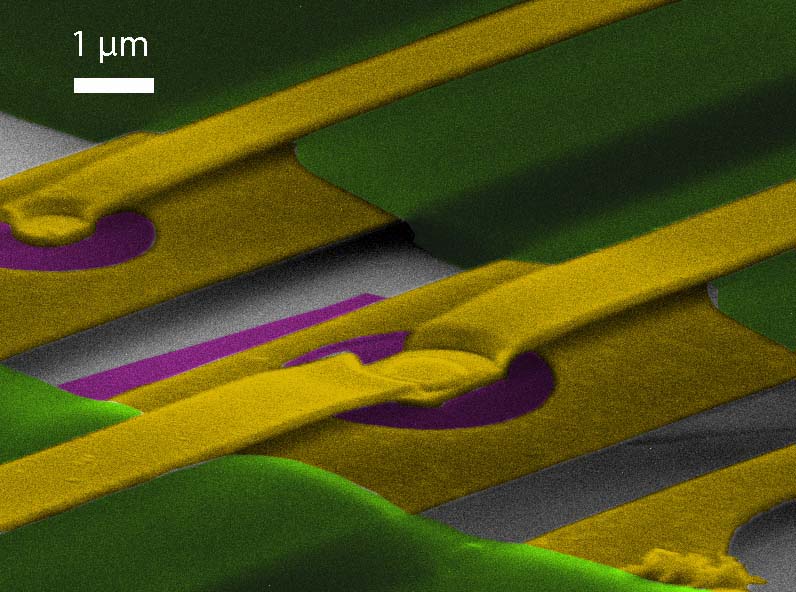}
\end{center}
\caption[]{An SEM micrograph of the measured samples: sample S1 on the left and S2 in the center. In the picture, graphene is marked using purple colour, Au/Cr contacts are in yellow, LOR in green, and gray background denotes the Si-substrate.}
\label{SEM}
\end{figure}

Two suspended graphene Corbino samples of similar characteristics were measured: S1 and S2, both depicted in Fig. \ref{SEM}. S1 had inner and outer diameters of $D_i$ = 0.8 $\mu$m and $D_o$ = 3.2 $\mu$m. In turn, S2 had inner and outer diameters of $D_i$ = 1.5 $\mu$m and $D_o$ = 3.8 $\mu$m. Before the actual measurements, the samples were current annealed using voltages 1.6 V (S1) and 2 V (S2), which resulted in high mobility in both samples. The Dirac point resided at $V_{gate} \approx -2$ V and the electron side mobility exceeded $10^5$ $\text{cm}^2/(\text{Vs})$ at $8\cdot 10^{10}$ $\text{cm}^{-2}$ for both samples.

The samples were mounted in a BlueFors LD-400 dilution refrigerator with base temperature of 10 mK. The cryostat was equipped with a superconducting magnet producing up to 9 T magnetic field perpendicular to the graphene surface. The DC-conductance measurements were conducted using RC-filtered ($R = $ 450 $\Omega$, $C = $ 30 nF) twisted pair lines. At high fields, the sample resistance in series with $RC$-filters remained high even at high bias voltages, which resulted in long $RC$ time constants in the system (e.g. $\tau = R_{sample}C$ = 3 s, when $R_{sample}$ = 100 M$\Omega$). Consequently, a pause time of 30 s was used for each bias point in the $IV$-sweeps.

The IV-measurements were carried out by applying a voltage bias at the top of the cryostat and measuring the current through the device using a transimpedance amplifier (Stanford Research Systems SR570). Thus, the IV data consisted of sets DC-voltages and corresponding currents through the Corbino disk. The current noise level in the measurements was $10^{-13}$ A. The charge carrier density was tuned by applying a voltage to the back gate. More specifically, the gate voltage was set to $V_{gate}$ = -2 V in order to stay in the middle of the 0th Landau level at the filling factor $\nu$ = 0.

%

\end{document}